# Algèbre OLAP et langage graphique

## F. Ravat, O. Teste, G. Zurfluh

*IRIT (SIG/ED)*
*118, route de Narbonne*
*31062 Toulouse cedex 04*

*{ravat, teste, zurfluh}@irit.fr*

RÉSUMÉ. *Cet article se situe dans le cadre des systèmes décisionnels reposant sur une modélisation multidimensionnelle. Le modèle conceptuel proposé représente les données sous forme de constellations (multi-faits) comportant des dimensions à hiérarchies multiples. Ce modèle restitue les données décisionnelles sous forme de tables multidimensionnelles. Ces tables servent de support à l'élaboration de notre algèbre. Cette algèbre, orientée décideur, intègre un noyau minimal fermé d'opérateurs OLAP complétés par des opérateurs avancés afin de faciliter l'expression des analyses décisionnelles complexes. Enfin, nous proposons un langage graphique complet au regard de l'algèbre. Grâce à ce langage, les décideurs peuvent exprimer simplement leurs analyses OLAP.*

ABSTRACT. *This article deals with OLAP systems based on multidimensional model. The conceptual model we provide, represents data through a constellation (multi-facts) composed of several multi-hierarchy dimensions. In this model, data are displayed through multidimensional tables. We define a query algebra handling these tables. This user oriented algebra is composed of a closure core of OLAP operators as soon as advanced operators dedicated to complex analysis. Finally, we specify a graphical OLAP language based on this algebra. This language facilitates analyses of decision makers.*

MOTS-CLÉS : *OLAP, algèbre multidimensionnelle, langage graphique*

# 1. Introduction

Les concepts liés aux systèmes d'aide à la décision tels que l'analyse en ligne des données (OLAP), les entrepôts de données (datawarehouse) et les Bases de Données Multidimensionnelles (BDM) n'ont pas toujours fait l'objet d'une formalisation précise. Pour le modèle multidimensionnel, des concepts et des systèmes existent sans fondement théorique stable (Marcel, 1998 ; Niemi, *et al.* 2003).

## *1.1. Contexte et problématique*

En l'absence d'un modèle consensuel pour les données multidimensionnelles, plusieurs propositions ont été présentées. La plupart de ces modèles repose sur un fait représentant un sujet d'analyse associé à des dimensions représentant les axes de l'analyse suivant lesquels différentes graduations hiérarchisées peuvent être adoptées (Kimball, 1996). Les états de l'art très complets de (Pedersen et al. 2001 ; Torlone, 2003) permettent de classer les travaux en deux catégories. La première intitulée « Modèle Cube » (Gray, *et al.* 1996 ; Gyssen, *et al.* 1997) propose de représenter les données sous forme de cube sans expliciter les différents composants d'un schéma multidimensionnel. La seconde catégorie, appelée « Modèle multidimensionnel », sémantiquement plus riche, permet d'expliciter précisément les différents composants d'un schéma multidimensionnel et notamment les différents niveaux d'agrégation des indicateurs (Agrawal, *et al.* 1997 ; Lehner, 1998 ; Pedersen, *et al.* 1999 ; Trujillo, *et al.* 2003 ; Abello, *et al.* 2003 ; Abello, *et al.* 2005).

Ces modèles servent de support aux algèbres de manipulation OLAP. Les premiers travaux ont adapté les opérateurs de l'algèbre relationnelle aux cubes de données (Gray, *et al.* 1996 ; Rafanelli, 2003). Pour pallier l'inadéquation de l'algèbre relationnelle au contexte OLAP, de nombreux travaux proposent de nouvelles opérations manipulant et générant des cubes (Cabibbo, *et al.* 1998 ; Pedersen, *et al.* 2001 ; Abello, *et al.* 2003 ; Franconi, Kamble, 2004). Ces propositions présentent l'inconvénient de ne pas se soucier de la manière dont sont restituées les données aux décideurs. Sachant qu'une visualisation sous forme de cubes en N dimensions (N>2) semble difficilement exploitable par les décideurs (Gyssen et al., 1997) et que ce type de présentation fait abstraction de la hiérarchisation des dimensions, nous avons proposé le concept de Table Multidimensionnelle (TM) avec une algèbre orientée décideurs (Ravat *et al.* 2001 ; Ravat *et al.* 2006).

La plupart de ces travaux n'intègre que quelques opérateurs de manipulation (forage –restitution des données en fonction des niveaux d'agrégation des dimensions–, rotation de dimensions et sélection). Ces propositions ne reposent pas toutes sur des définitions précises et formelles voire sur des modèles spécifiques tels que des modèles multidimensionnels à contraintes (Ravat, *et al.* 2006). A l'heure actuelle, il n'existe pas de consensus sur la définition d'un noyau minimum complet offrant une algèbre d'interrogation multidimensionnelle, à l'instar de l'algèbre relationnelle qui offre un support complet et reconnu.

*1.2. Proposition*

Dans ce cadre, notre objectif est de proposer un modèle de données multidimensionnelles servant de support à une algèbre et un langage graphique orientés décideurs pour **l'analyse et la visualisation** de données **OLAP.**

Pour ce faire, notre modèle doit répondre aux critères suivants : (i) faire abstraction des contraintes techniques de stockage de données afin de s'approcher de la vision des décideurs (Golfarelli, *et al*. 2002), (ii) faciliter les corrélations entre les différents sujets d'analyse et spécifier les fonctions d'agrégations compatibles pour chaque indicateur d'analyse (Sapia, *et al*. 1998 ; Abelló, *et al*. 2003 ; Trujillo, *et al*. 2003), (iii) intègrer la définition d'axes d'analyse à perspectives multiples d'une manière explicite, (iv) permettre de distinguer clairement les éléments de structures des valeurs, (v) proposer une visualisation facilement exploitable par les décideurs. En réponse à ces objectifs, nous proposons un modèle conceptuel de la catégorie "modèle multidimensionnel" intégrant une représentation multi-sujets d'analyse étudiés selon différents axes multi-hiérarchisés (Ravat, *et al*. 2006).

Ce modèle servira de support à l'élaboration d'une algèbre orientée utilisateurs permettant à un décideur d'effectuer des analyses décisionnelles complexes sur des schémas conceptuels et fournissant des TM (support plus adapté à la prise de décision que des cubes de données à N dimensions). En réponse aux différentes propositions d'algèbres multidimensionnelles, nous souhaitons fournir une algèbre OLAP fermée. Cette algèbre comprendra un noyau minimal complet d'opérateurs pouvant être combinés pour répondre aux besoins des décideurs ainsi qu'un ensemble d'opérateurs de second niveau permettant de simplifier des requêtes complexes. De plus, cette algèbre sert de support à un langage graphique destiné à des décideurs afin d'effectuer simplement et de manière incrémentale des requêtes complexes. Ce langage graphique doit être complet au regard de notre algèbre. Notre proposition repose sur une visualisation graphique du schéma conceptuel évitant ainsi de confondre l'abstraction conceptuelle multidimensionnelle et les structures logiques relationnelles (comme dans les logiciels tels que Oracle). Cette représentation permet une expression graphique et incrémentale des requêtes multidimensionnelles directement sur le schéma sans une connaissance à priori des structures R-OLAP et de la transcription SQL des requêtes (comme dans les éditeurs tels que Discoverer, Business Object…).

Cet article est organisé comme suit : la section 2 décrit les concepts et les formalismes de notre modèle multidimensionnel, la section 3 définit les opérateurs algébriques et la section 4 présente le langage graphique reposant sur l'algèbre.

**2. Modèle Multidimensionnel**

Le modèle que nous proposons repose sur une représentation multi-faits (multi-sujets d'analyse). Chacun de ces faits est analysé en fonction d'axes d'analyses (dimensions) muli-vues (multi-hiérarchisées).

## 2.1. *Dimension, hiérarchie et fait*

Une dimension est caractérisée par des attributs ; chaque attribut représente une façon de graduer l'axe d'analyse. Les différents attributs d'une dimension sont organisés au sein d'une ou plusieurs hiérarchies.

**Definition** : Une dimension $D_i$ est définie par $(N^{Di}, A^{Di}, H^{Di}, I^{Di})$ où $N^{Di}$ est son nom, $A^{Di} = \{a^{Di}_1, \ldots, a^{Di}_u\}$ représente les attributs, $H^{Di} = \{h^{Di}_1, \ldots, h^{Di}_y\}$ représente les hiérarchies et $I^{Di} = \{I^{Di}_1, I^{Di}_2, \ldots\}$ est l'ensemble des instances de $D_i$.

Une hiérarchie représente une perspective d'analyse précisant les niveaux de granularité en fonction desquels sont observées les valeurs analysées. Une hiérarchie $h^{Di}_x$ de $D_i$ est un chemin élémentaire acyclique débutant par l'attribut de plus forte granularité (All) et se terminant par celui de plus faible granularité (Id).

**Definition** : Une hiérarchie est définie par $(N^{Di}_x, Param^{Di}_x, Suppl^{Di}_x)$ où $N^{Di}_x$ est son nom, $Param^{Di}_x = <a^{Di}_{k0}, \ldots, a^{Di}_{kz}>$ est un ensemble ordonné décrivant la hiérarchie des attributs (chaque attribut, appelé paramètre, correspond à un niveau de granularité d'analyse), $a^{Di}_{k0}$=All et $a^{Di}_{kz}$=Id, et $Suppl^{Di}_x: Param^{Di}_x \to 2^{ADi-ParamDix}$ est une application spécifiant les attributs faibles qui complètent la sémantique des paramètres (chaque paramètre est associé à un ensemble d'attributs faibles).

Un fait est caractérisé par un ensemble d'indicateurs ou mesures généralement numériques et additives (ou semi-additives) (Kimball, 1996 ; Golfarelli, *et al*. 1998).

**Definition** : Un fait $F_j$ est défini par $(N^{Fj}, M^{Fj}, I^{Fj}, IStar^{Fj})$ où $N^{Fj}$ est le nom du fait, $M^{Fj} = \{m_1, \ldots, m_w\}$ est un ensemble de mesures, $I^{Fj} = \{I^{Fj}_1, I^{Fj}_2, \ldots\}$ est l'ensemble des instances de F, et $IStar^{Fj}$ est une fonction associant chaque instance de $I^{Fj}$ à une instance de chaque dimension liée au fait.

## 2.2. *Constellation*

Une constellation est la généralisation du modèle en étoile (Kimball, 1996) dans lesquels un seul fait est modélisé.

**Définition.** Une constellation CS est définie par $(N^{CS}, F^{CS}, D^{CS}, Star^{CS})$ où $N^{CS}$ est son nom, $F^{CS} = \{F_1, F_2, \ldots, F_p\}$ est un ensemble de faits, $D^{CS} = \{D_1, D_2, \ldots, D_q\}$ est un ensemble de dimensions et $Star^{CS} : F^{CS} \to 2^{DCS}$ est une fonction associant les faits aux dimensions afin de spécifier les sujets et les axes d'étude associés.

**Exemple.** Des décideurs désirent analyser les importations françaises avec les indicateurs suivants : le montant des importations (Montant), la quantité de produits importés (Quantité) et le nombre d'employés de la société française importatrice (NbEmployés). Les montants et les quantités appartiennent au même fait IMPORTATIONS et sont analysés en fonction du temps (DATES), des produits importés (PRODUITS), des sociétés françaises importatrices (SOCIETES) et des sociétés étrangères fournisseurs (FOURNISSEURS). Le nombre d'employés des sociétés caractérise un autre fait EFFECTIFS qui est observable suivant le temps (DATES) et les sociétés importatrices (SOCIETES).

La constellation associée à ces spécifications est modélisée comme suit : (i) $N^{CS}$ = 'SH_IMPORT', (ii) $F^{CS}$ = {IMPORTATIONS, EFFECTIFS}, (iii) $D^{CS}$ = {PRODUITS, DATES, SOCIETES, FOURNISSEURS}, (iv) $Star^{CS}$={EFFECTIFS→{DATES, SOCIETES}, IMPORTATIONS→{PRODUITS, DATES, SOCIETES, FOURNISSEURS}}. Pour représenter le schéma d'une constellation, nous adoptons des notations graphiques proches de (Golfarelli *et al*. 1998).

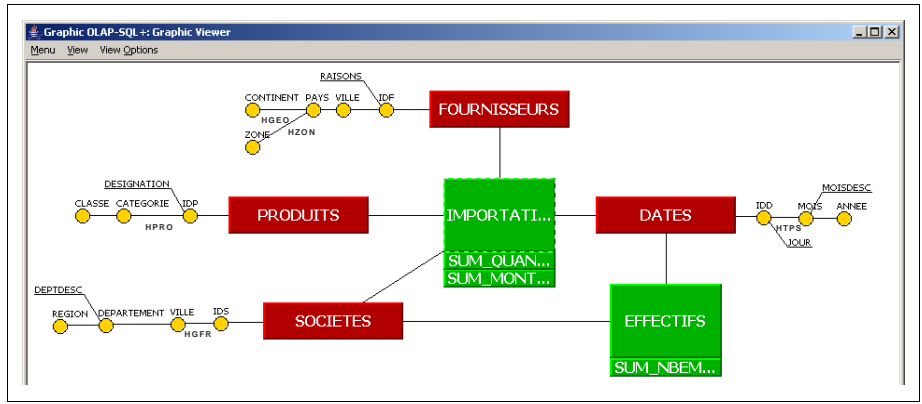

**Figure 1.** *Schéma d'une constellation.*

### 2.3. *Table multidimensionnelle*

La structure de visualisation que nous proposons repose sur un tableau à double entrées hiérarchisées, appelé Table Multidimensionnelle (TM). Une TM centre l'analyse sur un fait et facilite l'interprétation des données ; il s'agit d'une représentation très répandue (Gyssens *et al*. 1997) dans les outils décisionnels.

**Définition.** Une table multidimensionnelle T est définie par (S, L, C, R) où

- S = (F, {$f_1(m_1), f_2(m_2),...$}) représente le sujet d'analyse relatif au fait F et ses mesures observées {$m_1, m_2,...$} agrégées à l'aide de fonctions $f_1, f_2,...$,
- L = (DL, HL, <All, $p^{DL}_1, ..., p^{DL}_v$>) représente l'axe d'analyse en ligne de T au travers d'une dimension courante DL, d'une hiérarchie courante HL et d'une liste ordonnée de paramètres affichés <$p^{DL}_1, p^{DL}_2,...$> (le paramètre système All n'est pas affiché),
- C = (DC, HC, <All, $p^{DC}_1, ..., p^{DC}_w$>) représente l'axe d'analyse en colonne de la table T au travers d'une dimension courante DC, d'une hiérarchie courante HC et d'une liste ordonnée de paramètres affichés <$p^{DC}_1, p^{DC}_2,...$>,

– R = pred$_1$ ∧ pred$_2$ ∧… est le prédicat de restriction composé d'une conjonction de prédicats normalisés portant sur les dimensions et/ou F.

On pose dom($p^D_i$)=<$v_1$, $v_2$,…> le domaine de définition ordonné (valeurs) d'un paramètre $p^D_i$. Un exemple de TM est donné en figure 2.

## 3. Algèbre OLAP

Notre algèbre propose un ensemble d'opérateurs permettant à un décideur de manipuler les composants d'une constellation afin d'effectuer ses analyses. Notre algèbre repose sur 3 types d'opérateurs : un opérateur de construction produisant une TM à partir d'une BDM, un noyau minimum fermé d'opérateurs fondamentaux portant sur les TM et un ensemble d'opérateurs avancés facilitant les manipulations OLAP en offrant des fonctionnalités de plus haut niveau.

### 3.1. *Constructeur*

**Définition.** L'opération de construction est définie par

$$\text{DISPLAY}(N^{CS}, F, \{f_1(m_1), f_2(m_2),…\}, DL, HL, DC, HC) = T_{RES}$$

– $N^{CS}$ est le nom de la constellation,
– F est le fait analysé (sujet de l'analyse),
– {$f_1(m_1)$, $f_2(m_2)$,…} est un ensemble de mesures {$m_1$, $m_2$,…} du fait F agrégées à l'aide de fonctions $f_1, f_2,…$,
– DL est la dimension courante en ligne avec HL comme hiérarchie courante,
– DC est la dimension courante en colonne avec HC comme hiérarchie,
– $T_{RES}$=($S_{RES}$, $L_{RES}$, $C_{RES}$, $R_{RES}$) est la TM résultat où $S_{RES}$ =(F, {$f_1(m_1)$, $f_2(m_2)$,…}), $L_{RES}$=(DL, HL, <All, $p^{DL}_1$>) et $C_{RES}$=(DC, HC, <All, $p^{DC}_1$>) paramètres de plus haute granularité de HL et HC et $R_{RES}$=true.

**Exemple.** Nous considérons le schéma en constellation de la figure 1. Un décideur souhaite afficher la somme des montants importés par fournisseurs et par dates d'importations. La figure 2 présente le résultat de l'expression algébrique suivante : DISPLAY('SH_IMPORT', Importations, {*SUM*(Montant) }, Fournisseurs, HGeo, Dates, HTps) = $T_{R1}$

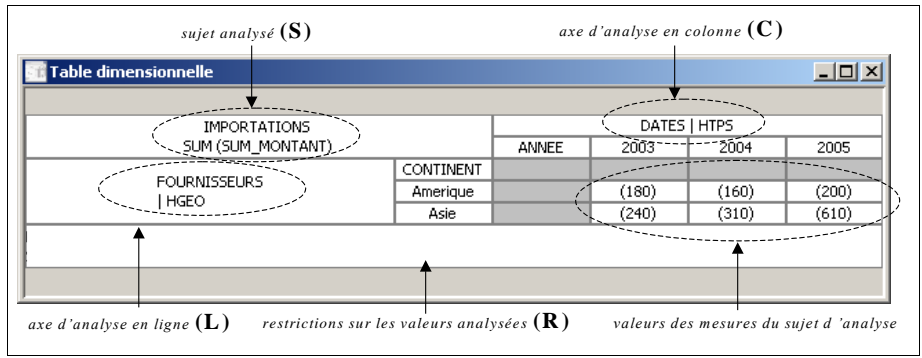

**Figure 2.** *TM résultat du DISPLAY.*

### 3.2. *Noyau minimum fermé*

Nous définissons un noyau minimum fermé d'opérateurs OLAP. Ces opérateurs s'appuient sur une TM et produisent en résultat une nouvelle TM ; cette propriété de fermeture de l'algèbre offre un cadre permettant d'élaborer des requêtes complexes par combinaisons d'opérations élémentaires (les définitions sont présentées en annexe). Le tableau 1 présente la syntaxe algébrique de ces opérateurs.

| Opérateur | Représentation symbolique |
|---|---|
| Rotation | DROTATE($T_{SRC}$, $D_{old}$, $D_{new}$, $H^{Dnew}_k$) = $T_{RES}$ |
| Forage vers le bas | DRILLDOWN($T_{SRC}$, D, $Att_{inf}$) = $T_{RES}$ |
| Forage vers le haut | ROLLUP($T_{SRC}$, D, $Att_{sup}$) = $T_{RES}$ |
| Imbrication | NEST($T_{SRC}$, D, Att, $D_{nested}$, $Att_{nested}$) = $T_{RES}$ |
| Sélection | SELECT($T_{SRC}$, pred) = $T_{RES}$ |
| Classement | SWITCH($T_{SRC}$, D, Att, $v_1$, $v_2$) = $T_{RES}$ |
| Agrégation | AGREGATE($T_{SRC}$, D, F(Att)) = $T_{RES}$ |
| Conversion d'un paramètre | PUSH($T_{SRC}$, D, Att) = $T_{RES}$ |
| Conversion d'une mesure | PULL($T_{SRC}$, $f_i(m_i)$, D) = $T_{RES}$ |
| Ajout/Suppression de mesures | ADDM($T_{SRC}$, $f_i(m_i)$)=$T_{RES}$   DELM($T_{SRC}$, $f_i(m_i)$)=$T_{RES}$ |

**Tableau 1.** *Noyau minimum d'opérateurs.*

Ces opérateurs sont les suivants :

– La rotation (DROTATE) permet, au sein d'une TM, soit de changer un axe d'analyse par un autre, soit de changer la hiérarchie sur un même axe.
– Les forages vers le bas ou vers le haut (DRILL-DOWN ou ROLL-UP) permettent au décideur d'analyser les données de manières plus ou moins

détaillées en modifiant les différents niveaux de graduation utilisés pour visualiser les données. Un niveau de graduation est représenté par un paramètre ou par des attributs faibles du paramètre spécifié (ou non).
- L'imbrication (NEST) permet d'intégrer, dans les dimensions d'une TM, les données provenant d'une ou plusieurs dimensions. Elle permet d'utiliser les paramètres de plusieurs dimensions dans l'espace 2D de la TM.
- L'opération de sélection (SELECT) permet de restreindre l'ensemble des valeurs affichées. Ces restrictions portent aussi bien sur les valeurs des attributs des dimensions que celles des mesures du fait.
- L'opération de classement (SWITCH) intervertit deux valeurs d'un attribut d'une dimension pour permettre l'ordonnancement des valeurs affichées.
- L'opération de calculs d'agrégats (AGREGATE) permet d'ajouter dans une TM des calculs agrégeant les lignes et/ou les colonnes. Cette opération correspond à l'opération Cube proposée par (Gray, *et al.* 1996).
- L'opération de conversion d'un paramètre en mesure (PUSH) permet de transformer un paramètre afin qu'il apparaisse dans la TM à l'utilisateur comme une mesure. Les valeurs du paramètre sont alors affichées dans les cellules contenant les valeurs des mesures analysées.
- L'opération de conversion d'une mesure en paramètre (PULL) transforme une mesure en paramètre de la TM. Les valeurs de la mesure sont affichées au niveau des entêtes en ligne ou colonne.
- Les opérations d'ajout (ADDM) et de suppression (DELM) de mesures permet de modifier l'ensemble des mesures analysées.

**Exemple.** Le décideur poursuit l'analyse précédente en focalisant son observation sur le montant et la moyenne des montants des importations en 2005 de produits électroniques. Il souhaite également affiner l'analyse en visualisant les montants plus finement, par pays d'origines de chaque fournisseur, tout en modifiant l'axe des colonnes pour observer les mesures par société importatrice. Pour ce faire, cette requête complexe est spécifiée par combinaison de plusieurs opérateurs élémentaires du noyau de l'algèbre :

- une sélection des PRODUITS 'Electronique' et des DATES en 2005,
- une opération de forage vers le bas sur l'axe des FOURNISSEURS,
- une opération de rotation des dimensions DATES et SOCIETES,
- une opération d'ajout de la mesure AVG(Montant).

L'expression algébrique suivante produit la TM décrite en figure 3.

DROTATE(ADDM(SELECT(DRILLDOWN($T_{R1}$, Fournisseurs, Pays), Produits.Classe = 'Electronique' $\wedge$ Dates.Annee = 2005), AVG(Montant)) , Fournisseurs, Societes, HGFr) = $T_{R2}$.

[Figure shows a "Table dimensionnelle" window with the following content:]

| IMPORTATIONS SUM (SUM_MONTANT), AVG (SUM_MONTANT) | | | SOCIETES \| HGFR | |
|---|---|---|---|---|
| | | | REGION | Midi-Pyrenees |
| FOURNISSEURS \| HGEO | CONTIN... | PAYS | | |
| | Amerique | Bresil | | (100, 100) |
| | | Etats-Unis | | (100, 100) |
| | Asie | Chine | | (230, 230) |
| | | Inde | | (160, 160) |
| | | Thailande | | (220, 220) |

PRODUITS.CLASSE = 'Electronique'
DATES.ANNEE = 2005

**Figure 3.** *TM résultat de la combinaison des opérateurs.*

### 3.3. *Opérateurs de second niveau*

Le noyau minimum de l'algèbre offre la possibilité de visualiser et d'effectuer des analyses plus ou moins complexes sur les données d'une constellation. Cependant, certaines analyses complexes nécessitent de nombreuses combinaisons d'opérateurs élémentaires du noyau. Afin d'améliorer le traitement des requêtes complexes, nous proposons un ensemble d'opérateurs de second niveau (construits par combinaison d'opérateurs du noyau minimum). L'intérêt de cette proposition est double : l'expression des analyses est réduite et les traitements systèmes correspondants aux opérations de second niveau peuvent être optimisés par rapport à la combinaison équivalente d'opérateurs du noyau.

Le tableau suivant présente les différents opérateurs de second niveau et l'expression représentant la combinaison équivalente d'opérateurs du noyau.

| Opérateur | Combinaison équivalente d'opérateurs du noyau |
|---|---|
| HROTATE($T_{SRC}$, D, $H^D_k$)=$T_{RES}$ | DROTATE($T_{SRC}$, D, D ,$H^D_k$)=$T_{RES}$ |
| PLOT($T_{SRC}$, D, Niv)=$T_{RES}$ | DRILLDOWN( ROLLUP($T_{SRC}$, D, All), D, Niv)=$T_{RES}$ |
| ORDER($T_{SRC}$, D, p, ord)=$T_{RES}$ ord∈{'asc', 'dsc'} | SWITCH(…(SWITCH($T_{SRC}$, D, p, $v_1$, $v_2$), …), D, p, $v_3$, $v_4$)=$T_{RES}$ |
| FROTATE($T_{SRC}$, $F_{new}$, {$f_1(m_1)$, $f_2(m_2)$,…})=$T_{RES}$ | History*($T_{SRC}$, DL, History* ($T_{SRC}$, DC, DISPLAY($N^{CS}$, $F_{new}$, {$f_1(m_1)$, $f_2(m_2)$,…}, DL, HL, DC, HC)))=$T_{RES}$ |
| UNSELECT($T_{SRC}$)=$T_{RES}$ | SELECT($T_{SRC}$, $F_{SRC}$.All='all' ∧ $D_{SRC\_1}$.All='all' ∧ … ∧ $D_{SRC\_q}$.All='all')=$T_{RES}$ |

**Tableau 2.** *Opérateurs de second niveau.*

(*) History($T_{old}$, obj, $T_{new}$)=$T_R$ représente l'historique des opérations qui ont été appliquées dans $T_{old}$ sur obj (fait ou dimension) et qui doivent s'appliquer à $T_{new}$.

Les opérateurs de second niveau offrent les fonctionnalités suivantes :

- L'opération de rotation de hiérarchies (HROTATE) consiste simplement à changer la hiérarchie courante d'une dimension ligne ou colonne.
- L'opération de projection d'un paramètre (PLOT) consiste à afficher les données suivant un paramètre quelconque de la dimension.
- L'opération d'ordonnancement (ORDER) croissant ou décroissant sur un paramètre consiste à ordonner les valeurs d'un paramètre.
- L'opération de rotation de faits (FROTATE), équivalente à l'opération Drill-Accross proposée par (Abello, *et al*. 2003), consiste à utiliser un nouveau fait dans la TM tout en conservant les caractéristiques des axes d'analyse courants. Cette opération n'est applicable que lorsque le nouveau fait partage au moins les deux dimensions courantes du fait de la TM initiale.
- L'opération de "désélection" (UNSELECT) consiste à annuler toutes les sélections sur les dimensions et le fait courants. Cette opération permet de construire une TM à partir de toutes les caractéristiques d'une TM initiale, mais en éliminant l'ensemble des restrictions sur les domaines de valeur.

**Exemple.** Le décideur modifie l'analyse de l'exemple précédent en focalisant son observation sur la somme et la moyenne des montants des importations de 2005 de produits électroniques. Il souhaite également affiner l'analyse en visualisant les montants plus finement, par pays d'origine de chaque fournisseur tout en modifiant l'axe des colonnes pour observer les mesures par sociétés importatrices. En réponse à ce besoin, l'expression algébrique suivante produit la TM décrite en figure 4.

PLOT(HROTATE(UNSELECT(DELM($T_{R2}$, AVG(Montant))), Fournisseurs, HZon), Societes, Ville) = $T_{R3}$.

| IMPORTATIONS SUM (SUM_MONTANT) | | SOCIETES \| HGFR | | | |
|---|---|---|---|---|---|
| | | VILLE | Bordeaux | Lyon | Toulouse |
| FOURNISSEURS \| HZON | ZONE | | | | |
| | E | | (140) | (100) | (1160) |
| | O | | (180) | (200) | (540) |

**Figure 4.** *TM résultat de la combinaison des opérateurs de second niveau.*

## 4. Langage Graphique GOLAP

### 4.1. *Visualisation d'un schéma multidimensionnel et construction d'une TM*

Notre objectif est d'offrir une vue globale des données analysables. Nous représentons une constellation par un graphe où chaque nœud est un fait ou une dimension et chaque arc représente un lien entre un fait et une dimension. Les nœuds se différencient grâce à des couleurs différentes : vert pour un fait et rouge pour une dimension. Cette représentation peut être développée en déployant les hiérarchies constituant chaque dimension. La visualisation développée du graphe introduit alors d'autres types de nœuds représentant les paramètres et les attributs faibles. Chaque dimension est alors constituée par un sous graphe de paramètres et d'attributs faibles hiérarchiquement liés (*cf*. figure 1).

La construction d'une TM s'effectue par manipulation directe et incrémentale du graphe. L'utilisateur sélectionne trois nœuds devant être un fait et deux dimensions ; le système assure la cohérence des sélections en rendant inaccessibles les nœuds invalides au fur et à mesure des sélections opérées par l'utilisateur.

**Exemple.** La TM de la figure 5 a été obtenue par manipulations graphiques et utilisation de menus contextuels. Le décideur a sélectionné le fait IMPORTATION, puis au travers d'un formulaire il a spécifié les mesures et les fonctions d'agrégation utilisées (∂). Ensuite, il a sélectionné incrémentalement les dimensions FOURNISSEURS (●) et DATES (÷) ; à chaque sélection un formulaire a permis de préciser la hiérarchie, les paramètres et/ou les attributs faibles affichés. Le décideur a validé la requête et le système produit en résultat une TM (≠).

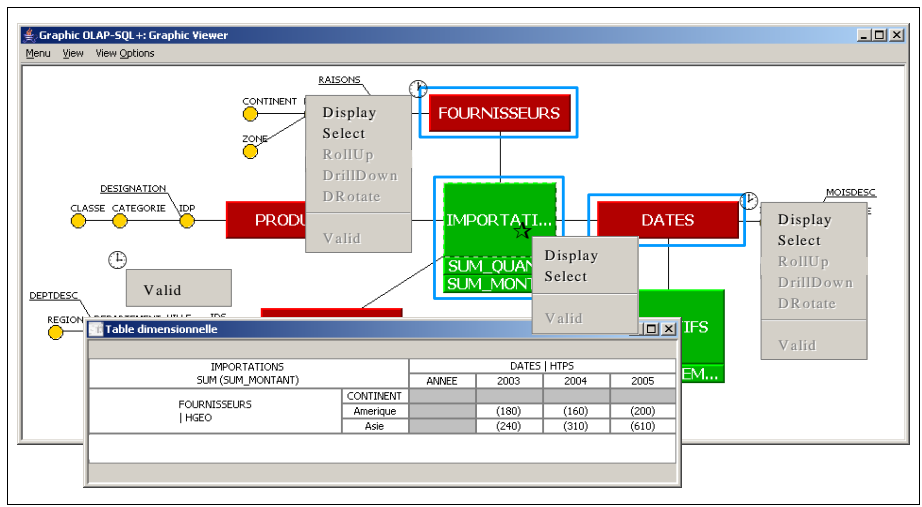

**Figure 5.** *Construction graphique d'une table multidimensionnelle.*

### 4.2. *Manipulations OLAP Graphiques*

Le système offre à un décideur deux manières d'effectuer des analyses décisionnelles. Un décideur peut agir graphiquement sur la constellation, mais il peut également appliquer certaines opérations directement sur les TM.

**Exemple.** Après construction de la TM précédente, le décideur poursuit son analyse. Plus précisément, ($\partial$) il affine son analyse en affichant les fournisseurs par pays d'origine, et (•) il focalise son analyse en sélectionnant la catégorie de produit intitulée 'Electronique'.

($\partial$) L'opération de forage permettant d'affiner la graduation sur l'axe d'analyse est exprimée en manipulant la dimension FOURNISSEURS soit sur le graphe, soit directement sur la TM obtenue précédemment (*cf.* figure 6).

(•) L'opération de sélection permettant de focaliser l'analyse uniquement sur les produits électroniques s'exprime en manipulant la dimension PRODUITS sur le graphe. Cette opération ne peut être définie qu'à partir du graphe puisque la dimension impliquée n'est pas disponible dans la TM.

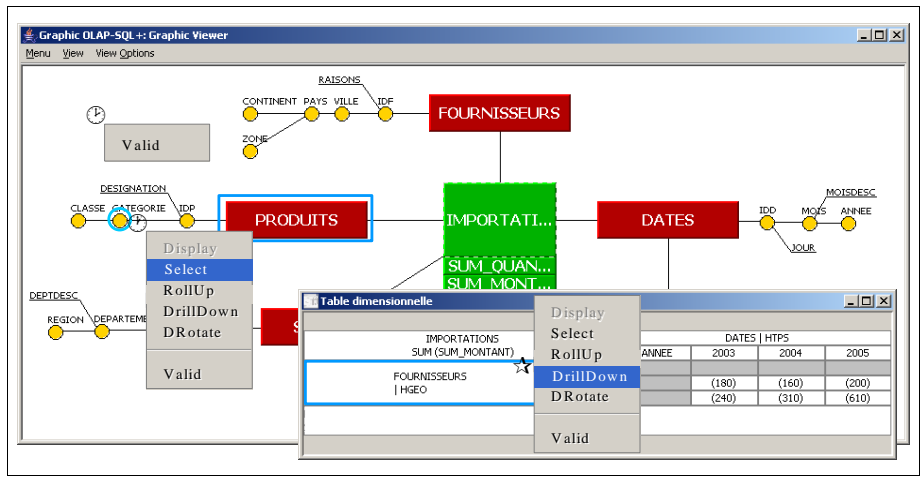

**Figure 6.** *Construction d'une requête.*

Toute opération de l'algèbre est exprimable au travers du langage GOLAP qui est donc complet au regard de notre algèbre multidimensionnelle.

### 5. Conclusion

Cet article se situe dans le cadre de l'analyse en ligne (OLAP) de données multidimensionnelles. Notre objectif est de pallier les manques des propositions actuelles en définissant : un modèle conceptuel multidimensionnel, une algèbre multidimensionnelle fermée orientée utilisateur, et un langage graphique complet reposant sur le fondement des opérateurs algébriques.

Notre modèle présente la caractéristique de permettre la définition de constellations multi-faits contenant des dimensions à hiérarchies multiples. Le modèle proposé permet de faire abstraction des aspects techniques tout en explicitant les composants de la structure d'une constellation.

Notre algèbre permet d'analyser les données d'un schéma multidimensionnel et de les visualiser au travers d'une TM. Cette algèbre, orientée décideurs, comprend trois types d'opérateurs multidimensionnels :

- un opérateur de construction de TM,
- un noyau minimal complet fermé d'opérateurs OLAP (rotations, forages, sélections, agrégations et transformations),
- un ensemble d'opérateurs avancés permettant de simplifier les expressions algébriques tout en permettant d'optimiser les traitements systèmes.

Cette algèbre sert de support à l'élaboration d'un langage graphique permettant à un décideur d'exprimer graphiquement une analyse de manière incrémentale. Ce langage est complet au regard de l'algèbre. Pour valider notre langage graphique, nous avons développé un prototype en java JDK1.5 au dessus du système de gestion de bases de données Oracle 10g ; ce prototype est composé de 100 classes. Il permet de définir et de manipuler une constellation R-OLAP ainsi que de visualiser et d'interroger les données multidimensionnelles au travers d'un graphe et de TM (les figures illustrant l'article sont extraites de ce prototype).

La majorité des bases de données multidimensionnelles étant implantées en relationnel, nous souhaitons compléter notre algèbre par la définition de processus de transformation des opérations multidimensionnelles en une combinaison optimale d'opérateurs relationnels sur des schémas R-OLAP normalisés (flocons) ou dénormalisés (étoiles et pré-agrégats).

## 6. Références bibliographiques

## 7. Annexe : Définition du noyau minimum fermé

Cette annexe définit les opérateurs de l'algèbre multidimensionnelle. Chaque opérateur prend une TM en entrée, notée $T_{SRC}$, et produit une TM, notée $T_{RES}$.

- $T_{SRC}=(S_{SRC}, L_{SRC}, C_{SRC}, R_{SRC})$ est la TM initiale.
- $T_{RES}=(S_{RES}, L_{RES}, C_{RES}, R_{RES})$ est la TM résultat. Par défaut, $S_{RES} = S_{SRC}$, $L_{RES} = L_{SRC}$, $C_{RES} = C_{SRC}$, $R_{RES} = R_{SRC}$ ; dans les définitions suivantes, nous spécifions uniquement les éléments de $T_{RES}$ modifiés.

Dans les définitions suivantes **Att** représente soit un paramètre p, soit un paramètre avec une liste des attributs faibles $p(a^D_{p1}, a^D_{p2},…)$, soit une liste d'attributs faibles $(a^D_{p1}, a^D_{p2},…)$ d'un paramètre p (non affiché dans le résultat).

---

**Rotation de dimensions** : $DROTATE(T_{SRC}, D_{old}, D_{new}, H^{Dnew}_k) = T_{RES}$

$D_{new}$ est la dimension qui remplace $D_{old} \in \{DC, DL\}$ dans la table résultat, $H^{Dnew}_k$ est la hiérarchie courante de $D_{new}$ (positionnée sur le paramètre de granularité maximale). $T_{RES}$ est la TM résultat telle que si $D_{old}=DL$ alors $L_{RES}=(D_{new}, H^{Dnew}_k, <All, p^{DL}_1>)$, si $D_{old}=DC$ alors $C_{RES}=(D_{new}, H^{Dnew}_k, <All, p^{DC}_1>)$.

---

**Forage vers le bas** : $DRILLDOWN(T_{SRC}, D, Att_{inf}) = T_{RES}$

$Att_{inf}$ représente un attribut inférieur dans la hiérarchie courante de la dimension D sur laquelle s'opère le forage. Les niveaux de graduation intermédiaires entre la graduation inférieure de $T_{SRC}$ et la nouvelle graduation ne sont pas pris en compte dans $T_{RES}$. $T_{RES}$ est la TM résultat telle que si D=DL alors $L_{RES}=(D, H^D_k, <All, p^{DL}_1,…, p^{DL}_v, Att_{inf}>)$, si D=DC alors $C_{RES}=(D, H^D_k, <All, p^{DC}_1,… p^{DC}_w, Att_{inf}>)$.

---

**Forage vers le haut** : $ROLLUP(T_{SRC}, D, Att_{sup}) = T_{RES}$

$Att_{sup}$ représente le niveau de graduation supérieur (dans la dimension D) utilisé dans $T_{RES}$, les graduations inférieures présentes dans $T_{SRC}$ sont supprimées. $T_{RES}$ est la TM résultat telle que si D=DL alors $L_{RES}=(D, H^D_k, <All, p^{DL}_1,…, Att_{sup}>)$, si D=DC alors $C_{RES}=(D, H^D_k, <All, p^{DC}_1,…, Att_{sup}>)$.

**Imbrication** : NEST($T_{SRC}$, D, Att, $D_{nested}$, $Att_{nested}$) = $T_{RES}$

Att est l'attribut de D au niveau duquel l'imbrication est effectuée et $Att_{nested}$ est l'attributs de $D_{nested}$ imbriqué. $T_{RES}$ est la TM résultat telle que si D=DL alors $L_{RES}$=(D, $H^D_k$, <All, $p^{DL}_1$,…, Att, $Att_{nested}$>), si D=DC alors $C_{RES}$=(D, $H^D_k$, <All, $p^{DC}_1$,…, Att, $Att_{nested}$>).

**Restriction** : SELECT($T_{SRC}$, pred) = $T_{RES}$

pred est un prédicat de sélection normalisé (conjonction de disjonctions) portant sur les dimensions et/ou le fait. $T_{RES}$ est la TM résultat où seul est remplacé le prédicat $R_{RES}$=pred.

**Permutation** : SWITCH($T_{SRC}$, D, Att, $v_1$, $v_2$) = $T_{RES}$

Att est l'attribut de la dimension D sur lequel s'effectue la permutation des valeurs $v_1$ et $v_2$ où dom(Att)=<…$v_1$,…$v_2$,…> dans $T_{SRC}$. $T_{RES}$ est la TM résultat où dom(Att)=<…$v_2$,…$v_1$,…>.

**Agrégation** : AGREGATE($T_{SRC}$, D, F(Att)) = $T_{RES}$

Att est l'attribut de la dimension D sur lequel s'applique la fonction d'agrégation F (sum, avg,…) avec dom(Att)=<$v_1$,…,$v_x$> dans $T_{SRC}$. $T_{RES}$ est la TM résultat où $\forall i \in [1..x]$, dom(Att) = <$v_1$, F($v_1$),…,$v_x$, F($v_x$)>. Chaque valeur initiale est complétée par la valeur représentant son agrégation.

**Destruction d'agrégation** : UNAGREGATE($T_{SRC}$) = $T_{RES}$

$T_{RES}$ est la TM résultat où toutes les valeurs agrégées sont éliminées.

**Conversion de paramètre en mesure** : PUSH($T_{SRC}$, D, Att) = $T_{RES}$

Att$\in H^D_k$ est l'attribut de la dimension D à convertir en mesure. $T_{RES}$ est la TM résultat où $S_{RES}$ = ($F_{SRC}$, {$f_1(m_1), f_2(m_2)$,…, Att}) avec Att$\notin H^D_k$.

**Conversion de mesure en paramètre** : PULL($T_{SRC}$, $f_i(m_i)$, D) = $T_{RES}$

$f_i(m_i)$ est une mesure du fait courant à convertir en paramètre de D. $T_{RES}$ est la TM résultat telle que si D=DL alors $L_{RES}$=(D, $H^D_k$, <All, $p^{DL}_1$,…, $f_i(m_i)$>), si D=DC alors $C_{RES}$=(D, $H^D_k$, <All, $p^{DC}_1$,…, $f_i(m_i)$>).

**Ajout de mesure** : ADDM($T_{SRC}$, $f_i(m_i)$) = $T_{RES}$

$f_i(m_i) \notin \{f_1(m_1),…,f_x(m_x)\}$ est une mesure à ajouter au fait courant de $T_{SRC}$. $T_{RES}$ est la TM résultat où $S_{RES}$ = ($F_{SRC}$, {$f_1(m_1),…,f_x(m_x), f_i(m_i)$}).

**Suppression de mesure** : DELM($T_{SRC}$, $f_i(m_i)$) = $T_{RES}$

$f_i(m_i) \in \{f_1(m_1),…,f_x(m_x)\}$ est une mesure à supprimer du fait courant. $T_{RES}$ est la TM résultat où $S_{RES}$=($F_{SRC}$, {$f_1(m_1),…,f_{i-1}(m_{i-1}), f_{i+1}(m_{i+1}),…,f_x(m_x)$}).